\documentclass[12pt]{article}    
\usepackage[dvips]{graphicx}

\begin{document}

\date{\today}
\title{On the Simulation of Extended TDHF Theory \footnote{
This work is supported in part by the U.S. DOE Grant No. DE-FG05-89ER40530.}}  
\author{Denis Lacroix$^{a)}$, Philippe Chomaz$^{a)}$ and Sakir Ayik$^{b)}$ \\
{\small $^{a)}$ {\it G.A.N.I.L., B.P. 5027, F-14021 Caen Cedex, France}} \\
{\small $^{b)}$ {\it Tennessee Technological University,
Cookeville TN38505, USA}}}
\maketitle

\begin{abstract}
{
 A novel method is presented for implementation of the extended 
mean-field theory incorporating two-body collisions. At a given time,
stochastic imaginary time propagation of occupied states are used to generate
a convenient basis. The quantal collision terms, including memory effects,
is then computed by a backward mean-field propagation of these single-particle
states. The method is illustrated in an exactly solvable model. 
Whereas the usual TDHF fails to
reproduce the long time evolution, a good agreement is found between the
extended TDHF and the exact solution.     
}
\end{abstract}

\vspace{5cm}



\section{Introduction}
Mean-field approximation is often employed to study static and dynamical
properties of many-body systems in different branches of
physics including atomic physics, condensed matter and nuclear physics \cite
{books}. The complex quantal many-body dynamics is reduced to an effective
one-body problem, in which particles move under an effective self-consistent
mean-field potential without experiencing correlations. Such an approach, 
which is usually referred to as Hartree-Fock
(HF) in static limit and Time-Dependent Hartree-Fock (TDHF) in dynamics, is
best suited at low energies at which binary collisions have little effects
on dynamics since scattering into unoccupied states is severely inhibited
due to Pauli blocking.
The HF and TDHF models have been very successful in describing many static
and dynamical properties in nuclear physics at low energies 
\cite{Vau2,Bon,Neg82,Dav84}. In
heavy-ion collisions around Fermi energy, dynamical evolution exhibits
strong dissipation and fluctuation properties. Mean-field alone is
inadequate to describe the collision dynamics at these energies, and it is
necessary to improve the one-body description beyond the mean-field
approximation by incorporating binary collisions due to short range
correlations into the equation of motion. This model is usually referred to
as the extended TDHF (ETDHF)\cite{ETDHF,Ayi80,Goek,Bal86}. 
A large amount of work has been done on the
formal development of the ETDHF, however, due to the numerical
complexity often linked with conceptual problems, the applications of
the theory on realistic situations remain a difficult problem, and only
a few approximate quantal calculations exists 
so far \cite{app_etdhf,Toh85,Tohayama,Giant}. Most of the
applications of this theory have been carried out in semi-classical
approximation, known as the Boltzmann-Uehling-Uhlenbeck (BUU) model
\cite{Ber88}. 
It has
been very successful for understanding a variety of features associated with
nuclear collisions at intermediate energies, including collective flow and
particle production. It is generally taken for granted that the test 
particle simulation
of the BUU model provides a good approximation for the average dynamics for
small quantal systems that we face with in heavy ion-collisions. The actual test
of the semi-classical models should be made by comparing the test particle
simulations with quantum transport calculations. Such a comparisons has been 
made in
the mean-field approximation in a recent work \cite{d1,RPA-Ber}, and 
demonstrated that
the expansion dynamics exhibits quantum effects which persists up to rather high
temperature $T\approx 5 MeV$. Furthermore, the quantal features may play even 
more
important role during the disassembly phase of the reaction. Therefore, it is 
off
great interest to develop quantal transport descriptions of heavy-ion dynamics.

In this paper, we propose a practical method for
obtaining numerical solutions of the ETDHF theory in its
quantal version, and illustrate the method in an exactly
solvable model\cite{Bormio}. 
We study a fermionic case but the method 
can be
applied to treat bosonic or even mixed systems.

\section{Extended Mean-Field Theory}

The basic idea of one body approaches is to project the dynamics onto 
the single-particle density matrix $\rho (t)$. In the 
ETDHF theory the evolution of $\rho (t)$ is determined by a transport
equation, 
\begin{eqnarray}
i\hbar \frac{\partial \rho }{\partial t}
-\left[ h\left[\rho \right],\rho \right] =K\left[ \rho \right].   
\label{eq:evol} 
\end{eqnarray}
The left hand side describes the evolution under the
effective mean-field Hamiltonian which may be related to the 
energy functional $E[\rho]$ as 
$h[\rho]=\partial E[\rho]/\partial \rho $ \cite{Ring}, and 
the right hand side represents a quantal binary collision term
\cite{Abe}. 
It is convenient to express the collision term
in the natural representation
$\left |\psi_{\lambda}(t)\right>$ which diagonalizes the single-particle 
density matrix 
\begin{eqnarray}
\rho (t)=\sum \left|\psi_{\lambda} (t)\right\rangle 
\;n_{\lambda}(t)\;\left\langle \psi_{\lambda} (t)\right|. 
\end{eqnarray}
Then, the matrix elements of the collision term are given by \cite{Tohayama} 
\begin{eqnarray}
\left<\psi_{\lambda}(t)|K\left[\rho\right]|\psi_{\lambda ^{\prime}}(t)\right> 
=-\frac{i}{\hbar }
\left( F_{\lambda ,\lambda ^{\prime}}(t)+
F_{\lambda ^{\prime },\lambda }^{*}(t)\right) 
\end{eqnarray}
with 
\begin{eqnarray}                                
\label{eq:COL}   
F_{\lambda ,\lambda ^{\prime }}(t) &=&\frac{1}{2}\int_{t_0}^{t}
dt^{\prime }\sum
\left. {\left\langle \lambda \delta^{\prime} |v_{12}
|\alpha \beta \right\rangle_A }\right|_{t}{\;}\left. 
\left\langle \alpha^{\prime} \beta^{\prime} | v_{12}|
\gamma\delta \right\rangle _{A}\right| _{t^{\prime }} \\
&&\left.\left(\rho_{\gamma \lambda ^{\prime }}
\rho_{\delta \delta^{\prime}}
\bar{\rho}_{\alpha \alpha^{\prime}}
\bar{\rho}_{\beta \beta^{\prime} }
-
\rho_{\alpha \alpha^{\prime}}
\rho_{\beta \beta^{\prime} }
\bar{\rho}_{\gamma \lambda^{\prime}}
\bar{\rho}_{\delta \delta^{\prime}}\right)\right|_{t^\prime}   \nonumber
\end{eqnarray}
where $\bar{\rho}_{\lambda\lambda^{\prime} }(t)=
\delta_{\lambda \lambda^{\prime}}-\rho_{\lambda\lambda^{\prime} }(t)$ and
$v_{12}$ denotes an effective residual interaction.
The matrix elements of the residual interaction are 
given as
\begin{eqnarray}
\left\langle \alpha \beta |v_{12}
|\lambda \delta \right\rangle_{A}| _{t^{\prime }}=
\left\langle \overline{\psi}_{\alpha} (t^\prime)\overline{\psi}_{\beta}(t^\prime)\right| 
\;v_{12}\;\left|\overline{\psi}_{\lambda}(t^\prime) 
\overline{\psi}_{\delta}(t^\prime)\right\rangle_{A} 
\end{eqnarray}
where $\left<\cdot \right>_{A}$ denotes the antisymmetrized matrix element,
and the states $\left|\overline{\psi}_{\lambda}(t^\prime)\right>$ 
are obtained by backward propagation of the natural states 
from time $t$ to $t'$  
by the mean-field
propagator,
$ \displaystyle \left|\overline{\psi}_{\lambda}(t^\prime)\right>=
U^\dagger(t,t^\prime)\left|\psi_{\lambda}(t)\right>$,
where the mean-field propagator is given by 
\begin{eqnarray}
U(t,t^\prime)=T \exp \left( -\frac{i}{\hbar }\int_{t^{\prime }}^{t}
h[\rho \left( s\right) ]ds\right) 
\end{eqnarray}
with $T$ as the time-ordering operator.
In the collision term (4), 
$\rho_{\lambda \lambda^\prime}(t^\prime)=
\left <\overline{\psi}_{\lambda}(t^\prime)\left|\rho(t^\prime)\right|
\overline{\psi}_{\lambda\prime}(t^\prime)\right> $ denotes the 
elements of density matrix at time $t^\prime$  in the basis of the backward 
propagated
states and it can be expressed as
\begin{eqnarray}
\rho_{\lambda \lambda^{\prime}}(t^\prime)=\sum 
\left <\overline{\psi}_{\lambda}(t^\prime)|\psi_{\alpha}(t^\prime)\right> 
n_{\alpha}(t^\prime)
\left <\psi_{\alpha}(t^\prime)|\overline{\psi}_{\lambda^{\prime}}(t^\prime)\right>
\end{eqnarray}
where $n_{\alpha}(t^\prime)$ denotes the occupation number of the natural state
$\left | \psi_{\alpha}(t^\prime)\right>$.

The collision term involves usually two
characteristic times: the correlation time $\tau _{cor}$ of the
matrix elements of the residual interactions and the relaxation time $\tau
_{rel}$ of the occupation numbers\cite{Weid80}. The expression of the 
collision term is valid when the correlation time is much smaller than the 
relaxation time, $\tau _{cor}\ll \tau _{rel}$, which is usually referred to
as the weak coupling limit.  
The time integration over the past history is evaluated 
over a time interval $t-t_{0}$ 
which should be sufficiently larger
than the correlation time  
$\tau_{cor}$. In this case, since during the correlation time, 
the natural states deviate only slightly
from the states propagated with the mean-field, we may approximate the
overlap as $\displaystyle \left<\overline{\psi}_{\lambda}(t^\prime)\left.
\right|\psi_{\lambda'}(t^\prime)\right> \simeq \delta_{\lambda \lambda'}$, 
and as a result, the collision term takes a simple form                
\begin{eqnarray}
F_{\lambda ,\lambda ^{\prime }}(t) & \simeq & \frac{1}{2}
\int_{t_0}^{t}dt^{\prime }\sum
\left. {\left\langle \lambda \delta |v_{12}
|\alpha \beta \right\rangle_A }\right|_{t}{\;}\left. 
{\left\langle \alpha \beta |v_{12}|
\lambda^{\prime}\delta \right\rangle_{A}}\right| _{t^{\prime }} \\
&&\left.\left(n_{\lambda ^{\prime }}
n_{\delta }
\bar{n}_{\alpha}
\bar{n}_{\beta }
-
n_{\alpha}
n_{\beta }
\bar{n}_{\lambda}
\bar{n}_{\delta}\right)\right|_{t^{\prime }}.   \nonumber
\end{eqnarray}
Furthermore, in accordance with the weak-coupling limit this expression may 
be further simplified by neglecting the evolution of the occupation numbers 
during the correlation time
by taking them as 
$n_{\lambda }(t^{\prime })\simeq n_{\lambda }(t)$.

\section{Numerical method}

For the purpose of numerical iteration, it is convenient to transform 
the transport eq.(1) into an integral form. 
Given the density matrix at a time $t$ in terms of the 
occupied states $\left|\psi_{\lambda}(t)\right>$ and associated 
occupation numbers $n_{\lambda }(t)$, 
its evolution 
during a time interval $\Delta t$ may 
be expressed according to 
\begin{eqnarray}
\label{eq:rho_t_plus_delta_t}
\rho (t+\Delta t)& =& U(t+\Delta t,t)\;\rho (t)\;U(t,t+\Delta t) \\
& &-\frac{i}{\hbar }\int_{t}^{t+\Delta t}dsU(t+\Delta t,s)K(s)U(s,t+\Delta t) 
\nonumber
\end{eqnarray}
where the first term represents the pure mean-field evolution and the second
term is the perturbation caused by the collision term during the time 
interval $\Delta t$. 
Single-particle states $\left | \overline{\psi}_{\lambda}(t+\Delta t)\right>$
obtained by propagating the natural states from time $t$ to $t+\Delta t$
according to (6), provide a useful basis to calculate 
density matrix over short time intervals.
In this representation, the elements of the density matrix at time 
$t+\Delta t$ may 
approximately be given by
\begin{eqnarray}
\rho_{\lambda\lambda^{\prime}}(t+\Delta t) \simeq
n_{\lambda }(t)\delta _{\lambda \lambda ^{\prime}}- 
\frac{1}{\hbar^2}\int_{t}^{t+\Delta t}ds 
[F_{\lambda\lambda^{\prime}}(s)+F_{\lambda^{\prime}\lambda}^{*}(s)]
\label{eq:rho}
\end{eqnarray}
where $F_{\lambda\lambda^{\prime}}(s)$ is computed according to 
Eq. (4) or its approximate
form (8)  assuming that over the time interval $\Delta t$
the natural states can be approximated by 
$\left | \overline{\psi}_{\lambda}(s)\right>$.
At time $t+\Delta t$ 
the new natural states $\left|\psi_{\lambda}(t+\Delta t)\right>$ 
and their 
occupation numbers $n_{\lambda}(t+\Delta t)$ 
are determined by diagonalizing the occupation matrix
$\rho_{\lambda\lambda^{\prime}}(t+\Delta t)$. Then, 
the iteration is continued into the next time step in a similar manner. 
The time interval $\Delta t$ of
the numerical iteration can be taken larger than 
the typical numerical time-step of the 
mean-field evolution, but it should be smaller than the correlation time
to insure at most one collision takes place during the interval.
When the collision term in eq.(9) is a small 
perturbation, 
a full diagonalization
of the density matrix is not needed, and the new states and the changes in
the occupation numbers can be calculated 
using perturbation
theory. 
However, to insure numerical stability, it is always possible
to include higher orders. 
The occupation numbers are determined by the diagonal elements of (11)
which can be transformed into a 
generalized master equation,
\begin{eqnarray}
\frac{d}{dt}n_{\lambda }(t)=\int_{t_0}^{t}dt^{\prime }\left\{ \bar{n_{\lambda}}
\left( t^{\prime }\right) {{\cal W}_{\lambda }^{+}}\left(t,t^{\prime
}\right) -n_{\lambda }\left( t^{\prime }\right) {{\cal W}_{\lambda }^{-}}
\left(t,t^{\prime }\right) \right\}   \label{eq:master}
\end{eqnarray}
where the gain ${\cal W}_{\lambda }^{+}$ and loss 
${\cal W}_{\lambda }^{-}$ kernels are given by 
\begin{eqnarray}
{\cal W}_{\lambda }^{+}(t,t^{\prime })=\frac{1}{\hbar ^{2}}
\sum {\ {\it Re}\left\{ \left\langle \lambda \delta |v_{12}|
\alpha \beta \right\rangle_A|_{t}
\left\langle \alpha \beta |v_{12}|
\lambda \delta \right\rangle_{A}|_{t^{\prime }}\right\} 
n_{\alpha }\left( t^{\prime }\right)
n_{\beta }\left( t^{\prime }\right) 
\bar{n}_{\delta }}\left( t^{\prime}\right) 
\end{eqnarray}
and 
\begin{eqnarray}
{\cal W}_{\lambda }^{-}(t,t^{\prime })=\frac{1}{\hbar ^{2}}
\sum {\ {\it Re}\left\{ \left\langle \lambda \delta |v_{12}|
\alpha \beta \right\rangle_A|_{t}
\left\langle \alpha \beta |v_{12}|
\lambda \delta \right\rangle_{A}|_{t^{\prime }}\right\} 
\bar{n}_{\alpha }\left( t^{\prime }\right) 
\bar{n}_{\beta }\left( t^{\prime }\right) 
n_{\delta }}\left( t^{\prime}\right). 
\end{eqnarray}
According to perturbation theory, 
when the states are non-degenerate,
the new natural states at time 
$t+\Delta t$ may be expressed as,
\begin{eqnarray}
&& \left|\psi_{\lambda} (t+\Delta t)\right\rangle  = 
\left|\overline{\psi}_{\lambda}(t+\Delta t)\right\rangle \\ 
&& -\frac{1}{\hbar ^{2}}\sum_{\lambda ^{\prime }\neq \lambda }
{\frac{1}{n_{\lambda}(t)-n_{\lambda^\prime}(t)}}
{\left|\overline{\psi}_{\lambda^{\prime}}(t+\Delta t)\right\rangle 
\int_{t}^{t+\Delta t}ds
\left\{ F_{\lambda \lambda ^{\prime}}(s)
+F_{\lambda ^{\prime }\lambda }^{*}(s)\right\} }.  \nonumber
\end{eqnarray}
This expression emphasizes importance of the non-diagonal terms in
the collision kernel which insure a proper transformation of the occupied
states. With the diagonalization procedure the collision term itself
decides the structure of the important single-particle states to be
populated. For the application presented in this paper, 
we have used a direct diagonalisation technics
instead of a perturbative approach to define the waves functions. 

The collision term $F_{\lambda\lambda^{\prime}}(t)$ 
involves, in addition to the
occupied single-particle states 
$\left|\psi _{\lambda} ^{occu}(t)\right\rangle $ 
which are known, a complete set of unoccupied states.
Specification of the most relevant set of intermediate states in a
reasonable manner has been the major difficulty in the implementation of the
ETDHF theory. Some effort has been made to generate the unoccupied
states by the TDHF evolution in the same manner as the occupied states
\cite{Tohayama}.
Since it is difficult to guess from the beginning of the reaction, 
the relevant unoccupied states, which are strongly coupled with
occupied states during a binary collision, may not be generated by a 
TDHF evolution.
Moreover, the unoccupied states generated in this manner may escape to 
continuum before having a chance to participate in a possible binary collision.

In this article, we propose an algorithm to construct the relevant unoccupied 
states at each stage of the iteration directly from the occupied states by
employing a stochastic imaginary time propagator method as follows. 
At the beginning 
of each time step 
$\Delta t$, 
a series of states is generated by repeated application
of  the imaginary time propagator 
$U^{(n)}_{\beta}=\exp\left[-\beta \left( h[\rho]+\delta h^{(n)} \right)\right]$
on the occupied states 
$U^{(n)}_{\beta} \left| \psi^{(n-1)}_{\lambda}(t)\right>  
=\left| \psi^{(n)}_{\lambda}(t)\right>$ with 
$\left| \psi^{(0)}_{\lambda}(t)\right>  
= \left| \psi^{occu}_{\lambda}(t)\right>  $.
The series of states obtained by application of the propagator
$\exp(-\beta h[\rho])$ generate a restricted subspace which has 
the same symmetry properties as the occupied states. On the other hand
relevant unoccupied states may have different symmetry properties, and
therefore, may lie outside of this subspace.
In order to ensure symmetry breaking and to perform a faster sampling of 
the relevant configuration\cite{Ots95}, a 
stochastic part $\delta h^{(n)}$ is added to the 
the mean-field Hamiltonian {
 at each application of} the propagator 
$U^{(n)}_{\beta}$.
In the case of open systems,
the continuum may also be treated by adding a density-dependent
constraining field 
into $\delta h^{(n)}$. 
These series of linearly independent states together with the occupied states 
are orthonormalized using
a Schmidt procedure after each application of
$U^{(n)}_{\beta}$. In this manner, it is possible to construct a suitable 
subspace of properly orthonormalized single-particle basis 
$\{\left| \psi_{\lambda} (t)\right\rangle \}$
including both occupied $\left|\psi _{\lambda} ^{occu}(t)\right\rangle $ 
and unoccupied states $\left|\psi_{ \lambda} ^{unoc}(t)\right\rangle $.
The subspace of the unoccupied states 
may be truncated 
further by diagonalizing the
mean-field Hamiltonian $h[\rho(t)]$
in this subspace
and by removing the states with energies larger than a maximum amount.
The high frequency components of the occupied states have only a minor effect
in the collision term, since they damp out very quickly in time and because 
of the time integration in eq. (4) and (6). The
construction with the help of the imaginary time propagator diminishes the
high frequency components, and at the same time, it generates a space of the 
unoccupied states in terms of linear combinations of predominantly 
the low frequency components of each occupied states.  
In practice, for an approximate treatment of the collision term, it may be 
sufficient to generate the 
relevant unoccupied states by a few applications of $U_{\beta}$ on the 
occupied states. The sensitivity of the results on the approximate
treatment can be checked by
enlarging the subspace of unoccupied states.
The unoccupied and occupied states are then evolved backward in time
using the mean-field propagator
order to properly take into account the past history in the collision term. 
We should note that, the high frequency components of the basis may
become important for increasing available energy, and the method may depend 
on the structure of the residual interaction. Therefore, the method should 
be tested carefully before applications to nuclear collision dynamics around 
Fermi energy.

\section{First Application}

{
 We illustrate the method in 
a one dimensional model problem of two
identical fermions coupled to a total spin projection $m_{S}=0$,
i.e. one with spin up and one with spin down,
which are moving in an external anharmonic potential and 
are interacting  
via
a short range two-body
force. The Hamiltonian of the system is given by 
\begin{eqnarray}
H=\sum_{i=1,2}{\ \left( \frac{p_{i}^{2}}{2m}+\frac{k}{2}{\left(x_{i}-x_0
\right)}^{2}+
\frac{k^{\prime }}{4}{\left(x_{i}-x_0\right)}^{4}\right) }+v_{12}
\end{eqnarray}
with $k=-0.04$ MeV/fm$^{2},$ $k^{\prime }=0.08$
MeV/fm$^{4}$, $x_0=9.3$ fm. In this schematic model, we avoid the problem of construction of a
G-matrix by employing a simple force without a hard core. This force mimics the
effective interaction usually employed in mean-field approaches, and also
simplifies the discussion on the role of the collision term. We should note that
the same effective interaction should be used in both the collision term and in
the mean-field part of the ETDHF equation. The two-body interaction is taken as 
\begin{eqnarray}
v_{12}=v_{0}\exp [-\frac{\left(
x_{1}-x_{2}\right) ^{2}}{2\sigma ^{2}}]
\end{eqnarray}
with $v_{0}=-4$ MeV and $\sigma =2$ fm. We consider the system
initially prepared in a constrained equilibrium state at a temperature T=5 MeV. 
The initial two-particle density matrix is, therefore, represented by 
\begin{eqnarray}
{\rm D}_{12}(t=0)=\sum_{i}{\ \left| \Psi _{i}\right\rangle \frac{\exp
[-E_{i}/T]}{Z}\left\langle \Psi _{i}\right| }  \label{eq:D}
\end{eqnarray}
where the initial two-particle states are determined by solving a
constrained Schroedinger equation, $(H-\lambda Q)\left| \Psi
_{i}\right\rangle =E_{i}\left| \Psi _{i}\right\rangle $ with 
$Q$ as a one-body constraining field, which is
taken as $Q = \sum_i \lambda(x_i) (x_i-x_0)^2$ 
with $\lambda = -0.24$ MeV/fm$^{2}$ for $x_i>x_0$ 
and $\lambda= -0.12$ MeV/fm$^{2}$ for $x_i<x_0$. 
We follow the 
exact Liouville von Neumann 
evolution of 
the two-particle density
matrix ${\rm D}_{12}(t)$ by evolving each two-particle states
$\left| \Psi _{i} (t)\right\rangle$
with the time-dependent Schroedinger equation.
We compare the exact evolution of the
single-particle density matrix $\rho _{1}(t)=tr_{2}{\rm D}_{12}(t)$ with the
approximate evolutions obtained in the TDHF and the ETDHF
descriptions, starting with the same initial conditions.
In the simulations, the numerical time-step is taken as $0.5$ fm/c whereas
the collision term is  
evaluated in larger time intervals of
$\Delta t = 6$ fm/c.  
In the calculation,
we employ a harmonic potential with a stochastic strength for $\delta h$
and 
two iterations of the imaginary time propagator appeared to be 
sufficient to 
generate the relevant unoccupied states. Figure 1 shows 
the time evolution of the local single-particle density $\rho
(x,t)$ as a function of the position $x$. Until about $100$ fm/c, the TDHF
calculation shown by dashed lines provides a good approximation for the
exact evolution indicated by dotted lines, but
is not able to reproduce it for larger times. 
On the other hand, the 
ETDHF results indicated by solid lines are in good agreement with the exact
evolution even at large times. 
Figure 2 illustrates the evolution of the expectation value of the center 
of mass coordinate $<X(t)>,$ as a function of time. Also for this
observable, the ETDHF result (solid line) follows closely the exact
evolution (dotted line), whereas the TDHF calculation (dashed line) deviates
from the exact evolution at large times. 
Time evolution of the occupation numbers of time-dependent single-particle
states are plotted in figure 3. The 
ETDHF calculations shown by dashed lines are very close to the exact
evolution 
indicated by open circles, which are   
obtained by a direct diagonalization of 
$\rho _{1}(t)=tr_{2}{\rm D}_{12}(t)$. 
In the pure TDHF approach, the
occupation numbers remain constant and equal to their initial values.

\section{conclusion}

Even though the formal development of the ETDHF theory was available
for some time, only a few applications in some simplified model problems have
been carried out so far. The major problem of the numerical implementation
originates from the difficulties for a realistic treatment of the collision
term in a suitable representation. In order to overcome this difficulty, we
propose a possible method in which the most important unoccupied states,
those strongly coupled with the occupied states through the collision
term, are calculated dynamically at each time step  
and then evolved backward in time using the mean-field propagator
in order to treat 
the past history in
the collision term.
We illustrate the method in an exactly solvable one-dimensional system of
two-particles, and find that the description of the  
one-body density matrix
in the ETDHF is in good agreement with the exact
evolution. In realistic applications  
to nuclear collision dynamics, the method 
may be implemented using the existing TDHF codes since it requires only 
standards calculation of the residual interaction
matrix elements and usual mean-field propagation in real and imaginary time.
In these applications, the computational 
effort is not small,
and may require a factor of about 20-100 more 
computation time than the corresponding TDHF simulation,   
but becomes manageable with high computational power of present day
computer technology.
In fact, a first step in this direction has been recently taken
\cite{d2}. In this work, collective vibrations at finite temperature
have been investigated in the small amplitude limit of the ETDHF theory,
including the collisional term. Such a calculation requires the
computation 
of many matrix element as in the full ETDHF description. The work for 
simulations of the full theory for large amplitude collective motion is 
currently in progress. 
We should also note that, the ETDHF theory is relevant to
not only nuclear dynamics.
Therefore, the proposed method may provide a useful tool for
describing dissipation and fluctuation phenomena in other {
quantal systems}.

\vspace{2cm}

One of us (S. A.) gratefully acknowledges GANIL Laboratory for a
partial support and warm hospitality extended to him during his visits
to Caen. Two of us (D. L. and Ph. C.) thank the Tennessee Technological
University for a partial support and warm hospitality during their visit.
We thank J.-P. Blaizot, A. Bulgac, 
M. Colonna, J. Randrup  
and E. Suraud for their useful remarks.
This work is supported in part by the US DOE Grant No.
DE-FG05-89ER40530.

\newpage

\begin{figure}[tbph]
\begin{center}
\includegraphics*[height=16cm,width=13cm]{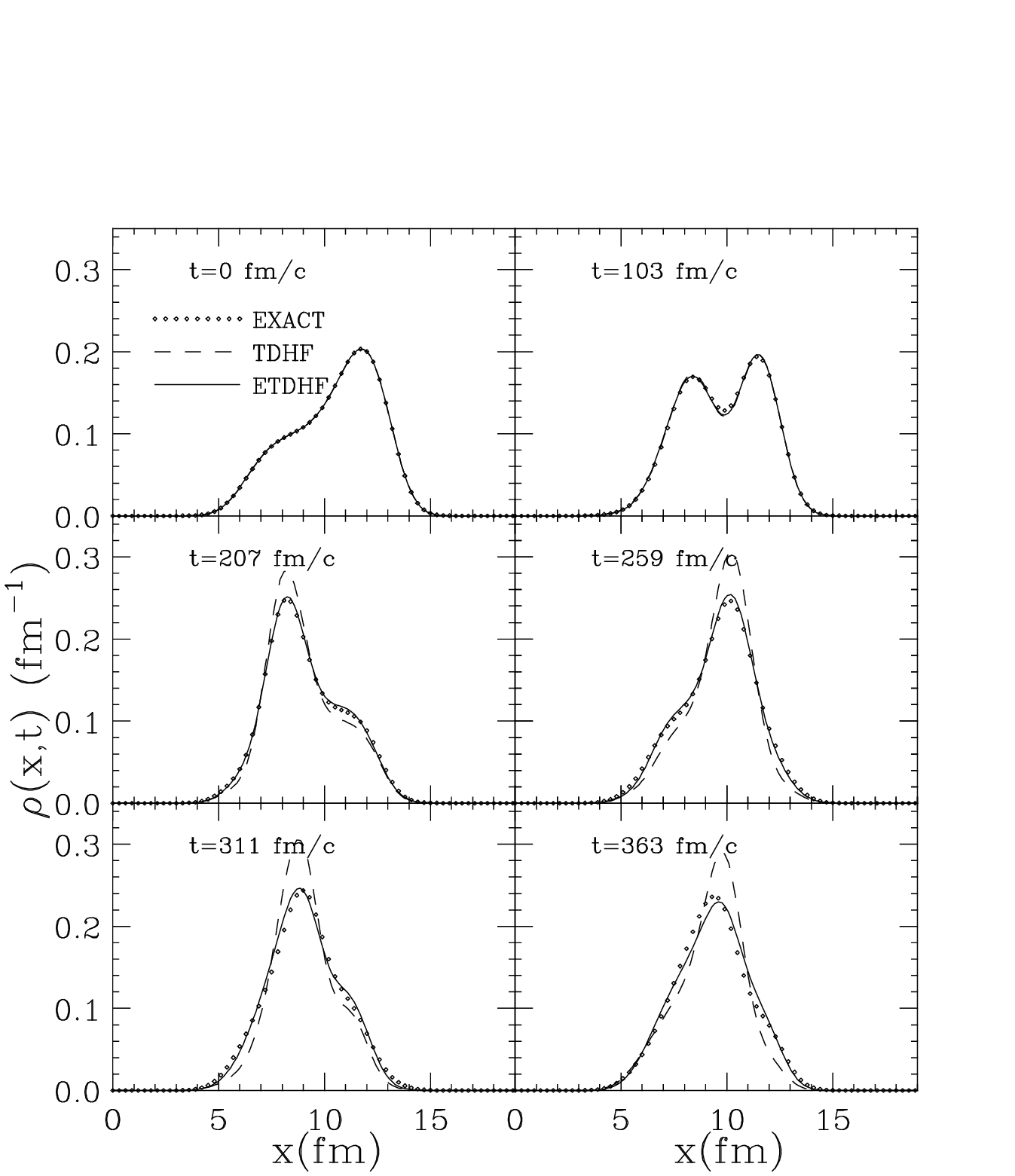}
\end{center}
\caption{Evolution of the local density. The exact calculations, the TDHF and
the ETDHF results are shown by circles, dashed lines and solid lines, 
respectively.}
\label{fig:5}
\end{figure}

\newpage

\begin{figure}[tbph]
\begin{center}
\includegraphics*[height=10cm,width=14cm]{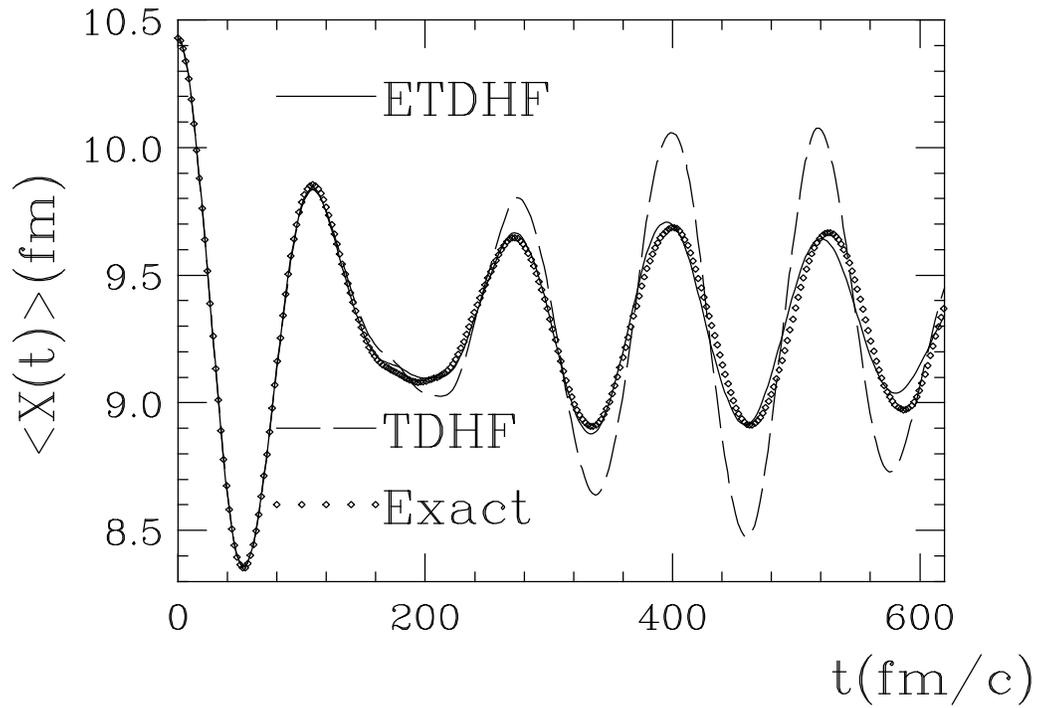}
\end{center}
\caption{Expectation value of the center of mass as a function of time.
The exact calculation, the TDHF and the ETDHF results are shown 
by circles, dashed line and solid line, respectively.}
\label{fig:6}
\end{figure}

\newpage

\begin{figure}[tbph]
\begin{center}
\includegraphics*[height=10cm,width=12cm]{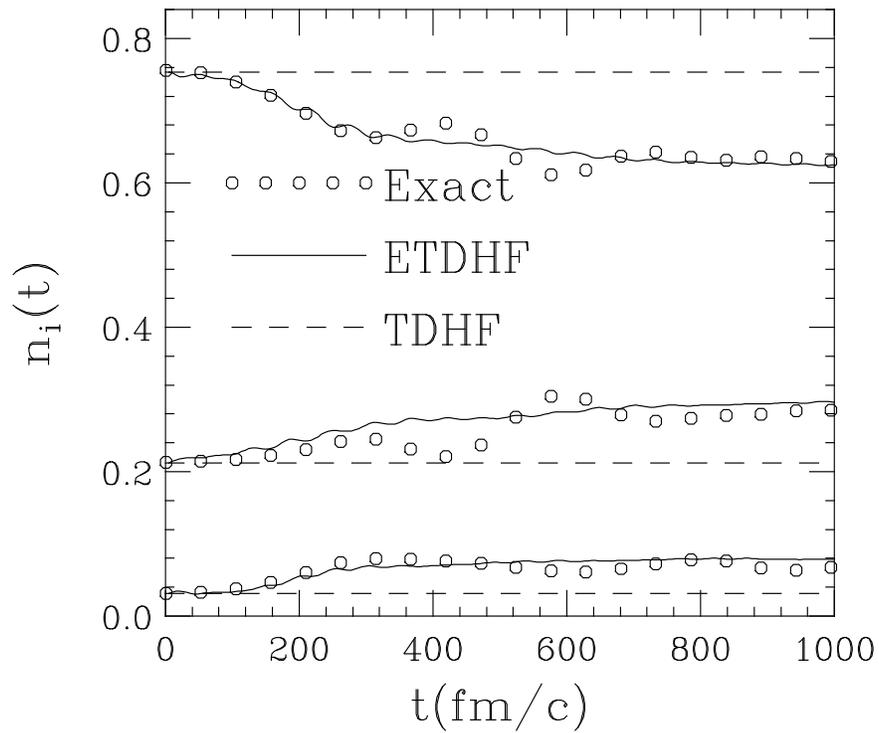}
\end{center}
\caption{Occupation numbers as a function of time. The exact calculations, the
TDHF and 
the ETDHF results are shown respectively by circles, dashed and solid
lines.}
\label{fig:7}
\end{figure}

\end{document}